\documentclass[letter]{aa}
\usepackage{graphicx}
\usepackage{txfonts}
\usepackage{natbib,graphicx,ulem}
\begin{document}
\title{Discovery of a new M32-like ``Compact Elliptical'' galaxy in the
halo of the Abell 496 cD galaxy \thanks{Based on observations obtained
at the Canada-France-Hawaii Telescope (program 03BF12) which is
operated by the National Research Council of Canada, the Institut des
Sciences de l'Univers of the Centre National de la Recherche
Scientifique and the University of Hawaii. Also based on ESO VLT data
(program 074.A-0533), and on HST archive data (proposals 5121 and 8683). } }

   \author{I. Chilingarian\inst{1,2}
          \and
          V. Cayatte\inst{3}
         \and
          L. Chemin\inst{4}
          \and
          F. Durret\inst{5,1}
           \and
          T. F. Lagan\'a\inst{6}
          \and
          C. Adami\inst{7}
          \and
          E. Slezak\inst{8}
          }

   \offprints{Igor Chilingarian \email{igor.chilingarian@obspm.fr}}

   \institute{Observatoire de Paris-Meudon, LERMA, UMR~8112, 61 Av. de l'Observatoire, Paris, 75014, France
         \and
              Sternberg Astronomical Institute, Moscow State University, 13 Universitetski prospect, 119992, Moscow, Russia
         \and
             Observatoire de Paris-Meudon, LUTH, UMR~8102, 5 pl. Jules Janssen, Meudon, 92195, France
         \and
             Observatoire de Paris-Meudon, GEPI, UMR~8111, 5 pl. Jules Janssen, Meudon, 92195, France
         \and
             Institut d'Astrophysique de Paris, CNRS, UMR~7095, Universit\'e Pierre et Marie Curie, 98bis Bd Arago, 75014 Paris, France
         \and
            Instituto de Astronomia, Geof\'{\i}sica e C. Atmosf./USP, R. do Mat\~ao 1226, 05508-090 S\~ao Paulo/SP, Brazil 
         \and
             Laboratoire d'Astrophysique de Marseille, UMR~6110, Traverse du Siphon, 13012 Marseille, France
         \and
             Observatoire de la C\^ote d'Azur, Laboratoire Cassiop\'ee, UMR~6202, 
             BP 4229, 06304 Nice Cedex 4, France
             }

   \date{Received March 01, 2007; accepted March 13, 2007; in original form February 13, 2007}

   \authorrunning{Chilingarian et al.}
   \titlerunning{cE in the Abell 496 cluster}

\abstract
{}
{``Compact ellipticals'' are so rare that a search for M32 analogs
is needed to ensure the very existence of this class. }
{We report here the discovery of A496cE, a M32 twin in the cluster 
Abell~496, located in the halo of the central cD. }
{Based on CFHT and HST imaging we show that the light profile of
A496cE requires a two component fit: a S\'ersic bulge and an
exponential disc.  The spectrum of A496cE obtained with the ESO-VLT
FLAMES/Giraffe spectrograph can be fit by a stellar synthesis spectrum
dominated by old stars, with high values of $[\mbox{Mg/Fe}]$ and
velocity dispersion.}
{The capture of A496cE by the cD galaxy and tidal stripping of most of its
disc are briefly discussed.}


\maketitle
%

\section{Introduction}

Compact elliptical galaxies are known to be high surface brightness
low-luminosity objects like the prototype of this class, \object{M~32},
satellite of the Andromeda galaxy. Compared to dwarf ellipticals of same
absolute magnitude, the effective surface brightness of \object{M~32} is
about 100 times higher and its effective radius is 10 times smaller (Graham
2002). Presently, only five objects belonging to the class of ``compact
elliptical'' galaxies (cE) are known; the four other objects are
\object{NGC~4486B} (in the vicinity of \object{M~87}, the Virgo cluster cD),
\object{NGC~5846A} (a close satellite of the giant elliptical galaxy
\object{NGC~5846}), and two objects in the \object{Abell~1689} cluster
(Mieske et al. 2005). These extremely rare galaxies are thought to be
generated by tidal stripping of more massive galaxies (Nieto \& Prugniel
1987, Bekki et al. 2001, Choi et al. 2002), but how such a high stellar
surface density can be put in the central part of the galaxy is not entirely
clear. The structural parameters of \object{M~32} derived by Graham (2002)
from the light profile include a bulge and a low surface brightness disc,
indicating that the precursor could be an early type disc galaxy. The
evolution in the past of the prototypical cE is however still a matter of
debate (see Mieske et al. 2005) and several observational projects aimed at
searching for cE galaxies were conducted until now with no success
(Drinkwater \& Gregg 1998, Ziegler \& Bender 1998).

\begin{figure}
\centering
\includegraphics[width=8cm]{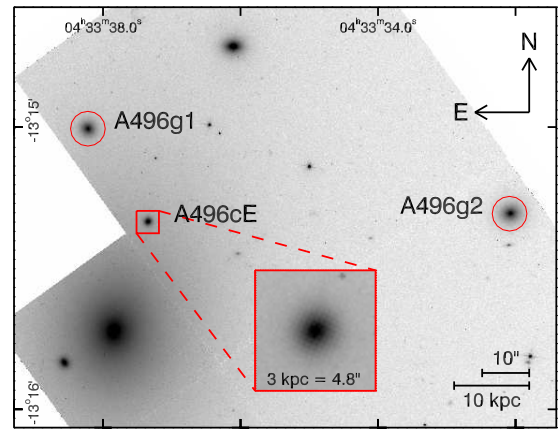}
\caption{
HST WFPC2 image of the central region of
Abell~496 (filter F702W). The objects discussed in this \textit{Letter} are
labeled.\label{figHST}}
\end{figure}

In this \textit{Letter} we report the discovery of the sixth ``compact
elliptical'' galaxy in the cluster \object{Abell~496}. We present results
based on the analysis of high-resolution spectroscopic data (R=7000 in the
5000--5800~\AA\ range) obtained with FLAMES/Giraffe (multi-object ``MEDUSA''
mode) at ESO VLT UT2 in the fall of 2004, combined with a photometric study
based on HST WFPC2 direct images in three filters: F555W, F702W, and F814W
available through the HST archive.  The discovery has been made from deep
ground-based u*~g'~r'~i' imaging conducted at CFHT with Megacam in the fall
of 2003 and dedicated to obtain morphological and structural properties of
dwarf early type galaxies in \object{Abell~496}. Detailed informations on
spectroscopic and photometric observations as well as data reduction will be
provided in a forthcoming paper discussing the whole sample of galaxies.

For the \object{Abell~496} cluster we adopt the distance modulus 35.70 and
spatial scale 0.627~kpc~arcsec$^{-1}$, assuming $H_0=73$~km~s$^{-1}$,
$\Omega_M=0.27$, and $\Omega_{\Lambda}=0.73$. All the photometric
measurements discussed in this {\textit Letter} are corrected for Galactic
absorption according to Schlegel et al. (1998), and redshift effect
($K$-correction) assuming an early-type galaxy spectrum. A cosmological
surface brightness dimming correction has been applied with the cluster
velocity, corrected for infall of the Local Group towards Virgo taken equal
to 9707~km~s$^{-1}$.

\section{Properties of the newly discovered object}

The galaxy is listed in HyperLeda\footnote{http://leda.univ-lyon1.fr/} as
\object{PGC~3084811} (Paturel et al. 2003). However, this name cannot be
resolved by other public astronomical databases (NED, SIMBAD), therefore we
adopt the IAU recommended name, \object{ACO496J043337.35-131520.2} and will
refer to it hereafter as ``A496cE''.

A496cE is located 22~arcsec (14~kpc in projected distance) from the cluster
centre; it appears in projection on the outer parts of the central cD galaxy
(see Fig.~\ref{figHST}). The background (i.e. the cD halo) has been
subtracted from the HST images using a multiscale wavelet analysis and
reconstruction technique described e.g. by Adami et al. 2005. The cD halo
surface brightness at the region where A496cE resides is about
24~mag~arcsec$^{-2}$ in the $B$ band and has a significant gradient,
therefore correct background estimation is crucial for the surface
photometry of A496cE. Elliptical isophotes with free central position,
ellipticity and position angle have been fitted to the images of A496cE
using the IRAF ELLIPSE task. On the r' image, the position angle changes
notably and the ellipticity rises from about 0.05 to 0.1 between 1.8 and 2.6
arcsec.

We have combined the HST light profiles for the inner part of the galaxy
($radius <$3.5~arcsec) with the deeper r' band CFHT photometry for the outer
parts ($radius >$3.5~arcsec), where the seeing (FWHM=0.8~arcsec) does not
play an important role. The empirical normalization factor for the r'
profile has been derived from the best match of the two profiles between 2
and 4 arcsec.

\begin{figure}
\centering
\includegraphics[width=7.5cm]{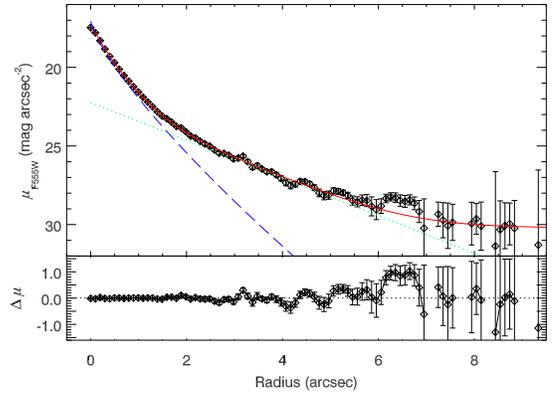}
\caption{Best fitting two-component S\'ersic+disc profile
overplotted on the composite light profile (F555 and r').
Bulge and disc components are shown
as dash-dotted blue and dashed green lines respectively.
\label{figfitprof}}
\end{figure}

We have fit the 1-dimensional surface brightness profile using both a
S\'ersic and a S\'ersic + exponential disc model (see Graham 2002 and Graham
\& Guzm\'an 2003) with a free constant background level within 10~arcsec
from the galaxy centre. As in the case of \object{M~32} (Graham 2002), a
single-component S\'ersic profile does not fit accurately the light profile
and an exponential disc is needed.  Fig.~\ref{figfitprof} and
Table~\ref{tabprofile} present the surface brightness profile of A496cE and
the best-fitting results of the photometric decomposition of the profile.

\begin{table}
\caption{
Global parameters (luminosity, radial velocity, central velocity dispersion)
and best-fitting parameters for the one-component (S\'ersic) or
two-component model (S\'ersic+disc) of the light profiles of A496cE, A496g1,
and A496g2 in $B$ band. Data for \object{M~32} are taken from van der Marel
et al. (1998), Graham (2002), and for NGC~4486B from Ferrarese et al. (2006)
and HyperLeda.\label{tabprofile}
}
\begin{tabular}{cccccc}
\hline
 & A496cE & A496g1 & A496g2 & M~32 & N4486B\\
\hline
$M_{B,tot}$    &-16.79 &-17.32 &-17.91 &-15.85 &-16.63 \\
$v$ km/s       & 9747$\pm$1     & 10286$\pm$3    &  9948$\pm$1    & -197$\pm$15 & 1557$\pm$35 \\
$\sigma_0$ km/s & 104$\pm$2      &   145$\pm$3    &    79$\pm$1    & 76$\pm$10   & 170$\pm$4 \\
\hline
$M_{B,b}$        &-16.60 &-17.32 &-17.91 &-15.34 &-16.63 \\
$R_{e,b}$ kpc    & 0.24  & 0.66  & 0.96  & 0.10  & 0.18  \\
$\mu_{0,b}$      & 17.37 & 18.70 & 17.79 & 16.31 & 14.00 \\
$n_{b}$          & 1.29  & 1.35  & 1.94  & 1.5   & 2.73  \\
$\mu_{e}$        & 19.82 & 21.30 & 21.65 & 19.23 & 19.58 \\
$<\mu>_{e}$      & 18.00 & 20.44 & 20.63 & 18.34 & 18.39 \\
\hline
$M_{B,d}$        &-14.80 & n/a   & n/a   &-14.78 & n/a   \\
$R_{e,d}$ kpc    & 0.92  & n/a   & n/a   & 0.84  & n/a   \\
$\mu_{0,d}$      & 22.55 & n/a   & n/a   & 22.28 & n/a   \\
\hline
\end{tabular}
\end{table}

An unsharp masking technique with elliptical blurring (Lisker et al. 2006),
shown to be very sensitive to the presence of possible embedded structures,
reveals no structures either on HST or on CFHT images.

We have built F555W-F702W and F555W-F814W colour maps using the Voronoi
adaptive binning technique (Cappellari \& Copin 2003) to achieve minimal
signal-to-noise ratios of 20 per bin in the F814W image and 40 in F702W.  No
colour gradient is detected.  However, this should not be considered as a
certain evidence for the uniform distribution of stellar population
parameters with radius. For instance, for \object{M~32} the effects of age
and metallicity gradients on the $V-I$ colour exactly compensate each other,
resulting in a flat colour profile (Rose et al. 2005).

We have analysed the spectrum of A496cE using the novel stellar population
fitting technique described in detail in Chilingarian (2006) and
Chilingarian et al. (2007). By fitting the spectrum with high-resolution
models of simple stellar populations (SSP) computed with PEGASE.HR (Le
Borgne et al. 2004), we are able to extract simultaneously stellar
kinematics: $v$, $\sigma$, and stellar population parameters such as
SSP-equivalent age and metallicity.  Although some template mismatch is seen
due to the supersolar $[\alpha/Fe]$ in A496cE, the quality of the fit is
rather good ($\chi^2/DOF = 1.6$).

\begin{figure}
\centering
\includegraphics[width=7.5cm]{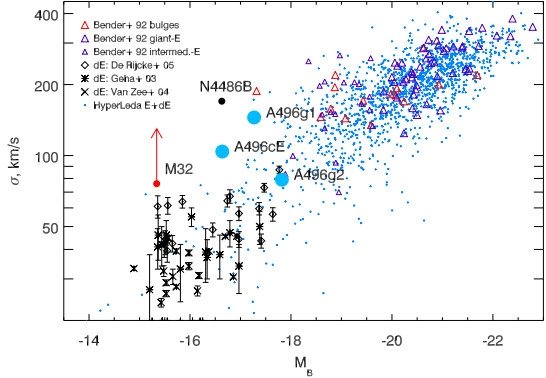}
\caption{Faber--Jackson relation for ellipticals and bulges of disc 
galaxies. Giant and intermediate-luminosity ellipticals are in blue, bulges
of disc galaxies in red, dwarf ellipticals and lenticulars in black. The
upper end of the arrow representing \object{M~32} corresponds to the HST
STIS measurements (Joseph et al. 2001), and the filled circle represents the
value obtained from earlier HST FOS data (van der Marel et al. 1998), which
was in agreement with more recent ground-based observations.
\label{figfjr}}
\end{figure}

We have obtained the $\alpha$/Fe abundance ratio of the stellar population
by measuring the following absorption line-strength indices (Worthey et al. 
1994): Mg$b$, Fe$_{5270}$, and Fe$_{5335}$ using the $\alpha$-enhanced
models by Thomas et al. (2003). The kinematical parameters are given in
Table~\ref{tabprofile} and those of the stellar populations in
Table~\ref{tabstpoppar}. A496cE has a rather high velocity dispersion for
its luminosity. It resides above the sequence of elliptical galaxies on the
Faber-Jackson (1976) relation, shown in Fig.~\ref{figfjr}, presenting a
compilation of data for dwarf (Geha et al. 2003; van Zee et al. 2004; De
Rijcke et al. 2005), intermediate luminosity, and giant elliptial galaxies
and bulges of bright lenticulars (Bender et al. 1992).  Due to the aperture
size of the FLAMES/Giraffe fibre our velocity dispersion for A496cE nearly
corresponds to the effective velocity dispersion and the central value is
probably significantly higher. The properties of A496cE, such as metallicity
and Mg/Fe ratio are similar to those of bulges of moderate-luminosity
lenticulars and spirals. The age and metallicity of A496cE correspond to a
stellar mass-to-light ratio $(M/L)_{*,B} = 19 \pm 2\ (M/L)_{\odot}$ (value
for the PEGASE.HR SSP computed with Salpeter IMF). Thus, the derived stellar
mass of A496cE is $(1.7 \pm 0.4)\cdot10^{10} M_{\odot}$.

\begin{table}
\caption{Stellar population parameters of A496cE, A496g1, and A496g2
compared to \object{M~32} and \object{NGC~4486B} (S\'anchez-Bl\'azquez et
al. 2006). Comma-separated values for \object{M~32} correspond to the
parameters at the core and at one effective radius as given by Rose et al.
(2005). \label{tabstpoppar}}
\centering
\begin{tabular}{cccc}
\hline
       & $t$, Gyr     &  $Z$, dex      &   $[\mbox{Mg/Fe}]$  \\
\hline
A496cE & 16.4$\pm$1.9 & -0.04$\pm$0.02 & 0.19$\pm$0.07  \\
A496g1 & 15.3$\pm$2.8 & -0.19$\pm$0.03 &  0.43$\pm$0.09 \\
A496g2 & 13.1$\pm$2.0 & -0.43$\pm$0.03 &  0.28$\pm$0.08 \\
M~32   & 4.0, 7.0     & 0.00, -0.25    &   -0.25, -0.08 \\
N4486B & 9.5          & 0.4            &  0.3           \\
\hline
\end{tabular}
\end{table}

\section{Discussion}
\begin{figure}
\centering
\includegraphics[width=7.8cm]{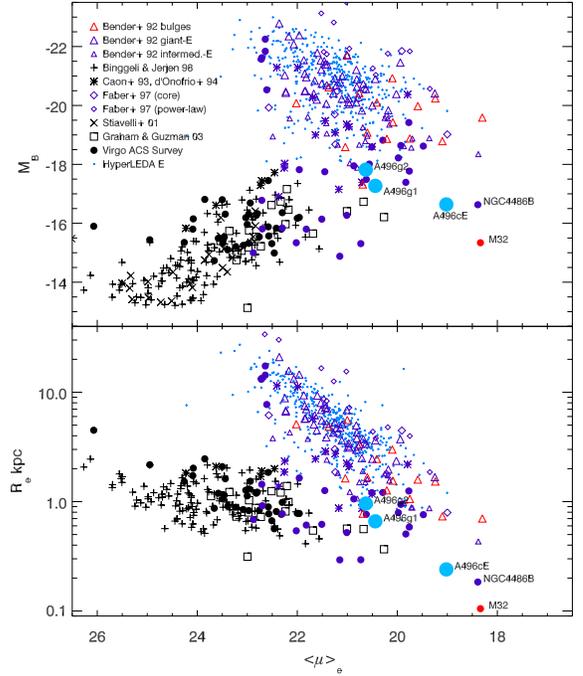}
\caption{Structural properties of elliptical galaxies and bulges.
Giant and intermediate-luminosity ellipticals, power-law and core galaxies
are shown in blue, bulges of disc galaxies in red, dwarf ellipticals and 
lenticulars in black. We keep the original morphological classification 
(E / dE) for the data points coming from the Virgo Cluster ACS Survey, 
Caon et al. (1993), and D'Onofrio et al. (1994), therefore they appear 
both in blue and black.}
\label{muplot}
\end{figure}

\subsection{Comparison with other E, dE, and cE galaxies}

The very small effective radius, about 250~pc, high mean surface brightness,
$\mu_{B e} = 19.60$~mag~arcsec$^{-2}$, and dwarf luminosity (${\rm
M_B}=-16.96$~mag) put the bulge of A496cE on the continuation of the
sequence of giant ellipticals and bulges in the Kormendy (Kormendy 1977) and
$M_B - \mu_{\mbox{eff}}$ diagrams toward small effective radii and fainter
luminosities. Fig.~\ref{muplot} presents the updated versions of Figs.~9a,g
from Graham \& Guzm\'an (2003): absolute $B$ magnitude and effective radius
$R_e$ versus mean $B$ surface brightness within effective radius $<\mu>_e$. 
The plot contains only data for elliptical galaxies and bulges of
lenticulars/spirals (where bulge/disc decomposition has been made in the
original sources), all integrated measurements for S galaxies have been
excluded. Data in computer-readable format for dE and E galaxies from
Binggeli \& Jerjen (1998), Caon et al. (1993), D'Onofrio et al. (1994),
Faber et al. (1997), Graham \& Guzm\'an (2003), Stiavelli et al. (2001) and
homogenization algorithms for these datasets have been kindly provided by
A.~Graham. We also included: photometric parameters of E and dE/dS0 galaxies
from the Virgo Cluster ACS Survey (Ferrarese et al.  2006), photometric data
on giant, intermediate elliptical galaxies and bulges of spirals and
lenticulars from Bender et al. (1992), photometric parameters of the
S\'ersic component of \object{M~32} (Graham 2002), and data for 430
elliptical galaxies from the HyperLeda database, with radial velocities
below 10000~km~s$^{-1}$ and brighter than $M_B=-18.0$~mag. All the
measurements are corrected for Galactic absorption, cosmological dimming,
K-correction, and converted into $B$ band according to Fukugita et al.
(1995) assuming an elliptical galaxy SED.

Reflecting structural properties of elliptical galaxies, giants and dwarfs
reside in different regions of these diagrams.  The sequence formed by giant
elliptical galaxies and bulges of spirals is clearly extended towards
smaller effective radii / higher surface brightnesses by \object{M~32} and
\object{NGC~4486B}. A496cE resides very close to \object{NGC~4486B} in both
plots. Thus, we can conclude that the structural properties of the A496cE
bulge resemble those of elliptical galaxies and bulges of spirals and
lenticulars.  However the bulges of \object{M~32} and A496cE lie in a region
where no other object is found, indicating that they differ from normal
bulges.

We have chosen S0 galaxies from the sample of Sil'chenko (2006), with
velocity dispersion and stellar population parameters coinciding with the
values for A496cE: \object{NGC~3098} ($\sigma=105$~km~s$^{-1}$,
$[\mbox{Fe/H}]=-0.2$, $t=10$~Gyr), and \object{NGC~4379}
($\sigma=108$~km~s$^{-1}$, $[\mbox{Fe/H}]=0.0$, $t=15$~Gyr). Their $B$
luminosities ($M_B=-18.9, -18.6$, HyperLeda) are about two magnitudes higher
than that of A496cE. Assuming a bulge-to-disc ratio of 1:1 by mass for
early-type galaxies, this implies that if the progenitor of A496cE was an
intermediate-luminosity disc galaxy, with its disc completely stripped by
harassment, it must have lost at least 2/3 of its bulge mass.

We now compare the properties of A496cE with two galaxies of our
spectroscopic sample in the WFPC2 field: \object{ACO496J043338.22-131500.7}
and \object{ACO496J043332.07-131518.1} (\object{PGC~93410}), hereafter
A496g1 and A496g2 respectively. The light profile of A496g1 is well fitted
with a single S\'ersic component law. We can also fit the A496g2 profile
with a single component model, though a faint disc in the external part may
also be present, as discussed in a forthcoming paper. For both objects the
effective radii are several times larger than for A496cE, while the
luminosities do not differ strongly. The metallicity and velocity dispersion
of A496g2 are not higher than expected for an object of such a luminosity,
but the $[\mbox{Mg/Fe}]$ ratio is very high, indicating a very short star
formation period (see e.g. Matteucci, 1994).  Gas might have been expelled
by ram pressure, efficient in the central parts of rich clusters (see e.g.
Abadi et al. 1999), so star formation was abruptly terminated.

For A496g1 the overall metallicity is $\sim$0.2~dex higher than expected for
its luminosity, but the $[\mbox{Mg/Fe}]$ ratio and velocity dispersion are
exceptionally high (see Fig.~\ref{figfjr}), so this galaxy is closer to an
intermediate-luminosity elliptical or S0 galaxy with $M_B\approx-19$~mag.
The old ages of the stellar populations of A496cE, A496g1, and A496g2 prove
that none of these objects merged with galaxies having young stellar
populations or star-formation during the last 10~Gyr, an indirect evidence
for their habitation in the centre of Abell~496 since that epoch.

\subsection{Possible origin of A496cE}

The presence of an outer exponential disc in A496cE argues that this galaxy
should not be considered as purely elliptical.  Lisker et al. (2007) have
proposed that only nucleated dwarf ellipticals follow the classical spheroid
picture and a large number of dEs in Virgo could be shaped like thick discs
and formed from mass loss from bigger infalling galaxies. However A496cE did
not follow the same evolutionary path as the three peculiar subclasses of
dwarfs described by Lisker et al., since its properties provide decisive
evidence that this object is quite unique.

Block et al. (2006) have simulated a head-on collision of Andromeda with a
low mass companion, now observed as \object{M~32}. Their numerical
simulations show that a fraction of M~32 gets stripped on a timescale of a
few $10^7$ years. Under certain circumstances if the disc galaxy is stripped
starting from its outer parts, the overall potential becomes shallower and
the bulge may shrink, leading to a smaller object.

A496cE is observed very close to the cluster centre. If it has a
short-period orbit partially immersed in the cD halo, consecutive passages
should lead to a relatively fast and efficient stripping. The old stellar
population of A496cE indicates that the main part of the disc was stripped
long ago.  But how could such an object survive for about 10~Gyr very close
to the cD galaxy without being accreted by it?

The two aspects that rule the processes of galaxy mergers are: (1) dynamical
friction, decreasing the orbit size and causing an accreting object to pass
closer and closer to the centre of the cluster; and (2) tidal forces, which
can totally disrupt an object if it passes sufficiently close to the 
cD galaxy. 

The bulge of A496cE is very compact and dense, thus it must be rather
resistant to tidal disruption.  Its exponential profile shows no evidence
for truncation beyond the tidal radius of A496cE $r_{tid} \sim $~2~kpc. We
assume a pericentral distance $d_p = 14$~kpc, i.e. that A496cE is passing
near its pericentre now with an orbital plane orthogonal to the line of
sight. This is possibly the case, because the radial velocity of A496cE
differs from that of the cD by only 100~km~s$^{-1}$.  Assuming a mass of
$\sim 10^{13} M_{\odot}$ for the cD, the pericentral velocity is $v \approx
2000$~km~s$^{-1}$. The dynamical friction is $\sim \frac{M^2 \rho}{v^2}$.
Given that A496cE is 4-6 times more massive (assuming the same dark matter
fraction) than M~32 ($\approx 3.0 \cdot 10^9 \mbox{M}_{\odot}$), the orbital
velocity 10 times higher, and halo densities are comparable, the dynamical
friction force is 3--4 times less efficient for A496cE than for M~32.
Therefore, there is a good chance for A496cE to survive for billions of
years in the central region of the cluster. Its progenitor must have lost a
large fraction of its mass during the first passage in order to decrease
significantly the dynamical friction effects. The high metallicity and
supersolar $[\mbox{Mg/Fe}]$ ratio are additional arguments for its massive
progenitor origin.

\begin{acknowledgements}
Special thanks to Alister Graham and Fran\c coise Combes for fruitful
discussions and to the anonymous referee for his/her fast and careful report
and useful suggestions. This \textit{Letter} was inspired by the panel
discussion "What is the relationship between compact ellipticals (such as
M32) and more massive ellipticals"\ at IAU Symposium 241.  We thank the CFHT
team for service observing and the Terapix team for reducing our Megacam
data. This paper has made use of the HST archive and of the NED and
HyperLeda data bases. IC acknowledges support provided by the ``HORIZON''
project, INTAS YS Fellowship (04-83-3618) and bilateral Russian-Flemmish
grant RFBR-05-02-19805-MF\_a.

\end{acknowledgements}

\end{document}